\documentclass[prd,unsortedaddress,superscriptaddress,a4paper,nofootinbib,%
nobalancelastpage,preprintnumbers,showpacs]{revtex4}

\usepackage{subfigure}
\usepackage[pdftex]{graphicx}
\usepackage{times}
\usepackage{amsmath,amssymb,amsbsy}
\begin{document}
\author{Cafer Ay} 
\affiliation{Laboratoire de Physique Subatomique et Cosmologie,\\
Universit\'e Joseph Fourier, CNRS-IN2P3, INPG\\
53, avenue des Martyrs, Grenoble, France}
\author{Jean-Marc Richard}
\email{jean-marc.richard@lpsc.in2p3.fr}
\affiliation{Laboratoire de Physique Subatomique et Cosmologie,\\
Universit\'e Joseph Fourier, CNRS-IN2P3, INPG\\
53, avenue des Martyrs, Grenoble, France}
\author{J. Hyam Rubinstein}
\affiliation{Department of Mathematics and Statistics,\\
University of Melbourne, Parkville 3010, Australia}
\date{\today}
\title{\sc Stability of asymmetric tetraquarks in the minimal-path linear potential}
\preprint{LPSC-0916}
\begin{abstract}
The linear potential binding a quark and an antiquark in mesons is generalized to baryons and multiquark configurations as the minimal length of flux tubes neutralizing the color, in units of the string tension.  
For tetraquark systems, i.e., two quarks and two antiquarks, this involves the two possible quark--antiquark pairings, and the Steiner tree linking the quarks to the antiquarks. 
A novel inequality for this potential demonstrates rigorously that within this model the tetraquark is stable in the limit of large quark-to-antiquark mass ratio.
\end{abstract}
\maketitle
The quark--antiquark  confinement in ordinary mesons is often described by a linear potential $V_2=r$, in units where the string tension is set to unity. For a given interquark separation $r$, it can be interpreted as the minimal gluon energy if the field is localized in a flux tube of constant section linking the quark to the antiquark.

\begin{figure}[hbtc]
\centering
\subfigure[\ Baryons]{\label{fig:bar-tetra-bar}
\includegraphics{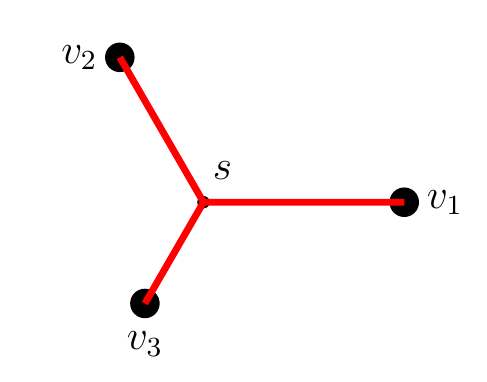}}
\hspace*{.5cm}
\subfigure[\ Tetraquarks: flip--flop]{\label{fig:bar-tetra-ff}
\includegraphics{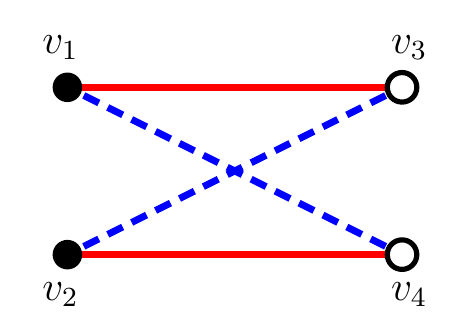}}
\hspace*{.5cm}
\subfigure[\ Tetraquarks: Steiner]{\label{fig:bar-tetra-Steiner}
\includegraphics{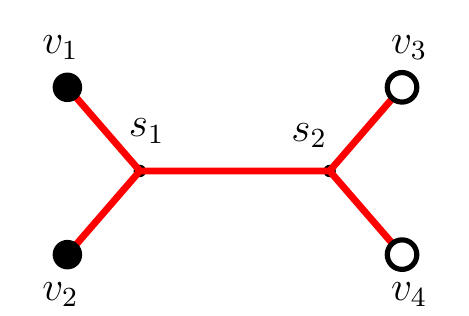}}
\caption{\label{fig:bar-tetra} Generalization of the linear quark--antiquark potential of mesons to baryons (left) and to tetraquarks, where the minimum is taken of the flip--flop (center) and Steiner tree (right) configurations.}
\end{figure}

The natural extension to describe the confinement of three quarks in a baryon is the so-called $Y$-shape potential
\begin{equation}\label{Y-shape}
V_3(v_1,v_2,v_3)=\min_s(d_1+d_2+d_3)~,
\end{equation}
where $d_i$ is the distance of the $i^\text{th}$ quark located at $v_i$ ($i=1,\,2,\,3$)  to a junction $s$ whose location is adjusted to minimize $V_3$. This potential has been proposed in Refs.~\cite{Artru:1974zn,Hasenfratz:1980ka,Dosch:1982ep,Bagan:1985zn,FabreDeLaRipelle:1988zr,Kuzmenko:2000rt,Takahashi:2000te}, among others. It has been used, e.g., in Refs.~\cite{Richard:1983mu,Carlson:1982xi} for studying the spectroscopy of baryons. See, also \cite{SilvestreBrac:2003sa}.
The optimization in (\ref{Y-shape}) corresponds to the well-known problem of Fermat and Torricelli to link three  points with the minimal network.  See Fig.~\ref{fig:bar-tetra-bar}.

We now turn to the  tetraquark systems $(Q,Q,\bar{q},\bar{q})$, with  the notation $(v_1,v_2,v_3,v_4)$ for the locations, and $(M,M,m,m)$ for the masses which will be used shortly. The potential is assumed to be (with $d_{ij}=\| v_i v_j\|$)
\begin{equation}\label{tetra-pot}\begin{aligned}
U&=\min \left\{ d_{13}+d_{24},d_{14}+d_{23},V_4 \right\}~,\\
\qquad
V_4&=\min_{s_1,s_2}\left(\, \|v_1s_1\| + \| v_2s_1\|+ \| s_1s_2\|+\| s_2v_3\|+\| s_2v_4\|\,\right )~.
\end{aligned}\end{equation}
The first two terms of $U$ describe the two possible quark--antiquark links, and their minimum is sometimes referred to as the ``flip--flop'' model, schematically pictured in Fig.~\ref{fig:bar-tetra-ff}. 
It was introduced by Lenz et al.~\cite{Lenz:1985jk}, who used, however, a quadratic instead of linear rise of the potential as a function of the distance. The last term, $V_4$, is represented in Fig.~\ref{fig:bar-tetra-Steiner} and corresponds to a connected flux tube. It is given by a Steiner tree, i.e., it is minimized by varying the location of the Steiner points $s_1$ and $s_2$.
The choice of this potential is inspired by Refs.~\cite{Chan:1978nk,Montanet:1980te,Dosch:1982ep,Martens:2006ac}, and has been discussed in the context of lattice QCD \cite{Okiharu:2004ve,Suganuma:2008ej}.

The four-body problem in quantum mechanics is notoriously difficult. For instance, Wheeler proposed in 1945 the existence of a  positronium molecule  $(\rm e^+,e^+,e^-,e^-)$ which is stable in the limit where internal annihilation is neglected, i.e., lies below its threshold for dissociation into two positronium atoms.  In 1946, Ore  published a four-body calculation of this system \cite{PhysRev.70.90} and concluded that his investigation ``counsels against the assumption that clusters of this (or even of higher) complexity can be formed''. However, in 1947, Hylleraas and the same Ore published an elegant analytic proof that this molecule is stable  \cite{PhysRev.71.493}. It has been discovered recently \cite{2007Natur.449..195C}.

Similarly, the above model (\ref{tetra-pot}), in its linear version, was considered by Carlson and Pandharipande, who entitled their paper \cite{Carlson:1991zt} ``Absence of exotics in the flux tube model'', i.e.,  did not find stable tetraquarks%
\footnote{The authors used a relativistic form of kinetic energy and considered also the possibility of short-range corrections, but this seemingly does not affect their conclusion.}.
 However,  Vijande et al.~\cite{Vijande:2007ix} used a more systematic variational expansion of the wave function and in their numerical solution of the four-body problem found a stable tetraquark ground state. Moreover, unlike \cite{Carlson:1991zt}, they considered the possibility of unequal masses, and found that stability  improves if the quarks are heavier (or lighter) than the antiquarks, in agreement with previous investigations (see, e.g., \cite{Vijande:2007ix} for Refs.).

It is thus desirable to check  whether this minimal-path model supports or not bound states. The present attempt is based on an upper bound on the potential, which leads to an \emph{exactly solvable} four-body Hamiltonian.

With the Jacobi vector coordinates
\begin{equation}\label{eq:Jacobi}
x=v_2-v_1~,\quad
y=v_4-v_3~,\quad
z=\frac{v_3+v_4-v_1-v_2}{2}~,
\end{equation}
and their conjugate momenta, the relative motion is described by the Hamiltonian
\begin{equation}\label{eq:H}
H=\frac{p_x^2}{M}+\frac{p_y^2}{m}+\frac{p_z^2}{4\mu}+ U(x,y,z)~,
\end{equation}
where $\mu$, given by $\mu^{-1}=m^{-1}+M^{-1}$, is the quark--antiquark reduced mass. Using the scaling properties of $H$, one can set $m=1$ without loss of generality.

The simplest bound on the potential $U$ is 
\begin{equation}\label{eq:boundU1}
U\le V_4\le \|x\|+\|y\|+\|z\|~,
\end{equation}
as the  tree with optimized Steiner points $s_1$ and $s_2$ is shorter  than if the junctions are set at the middles of the quark separation $v_1v_2$ and antiquark separation $v_3v_4$. This leads to a separable upper bound for the Hamiltonian
\begin{equation}\label{eq:boundH1}
H\le H'=\frac{p_x^2}{M}+ \|x\|+p_y^2+\|y\|+\frac{p_z^2}{4\mu}+\|z\|~.
\end{equation}
Now, the ground state $e_0$ of $p_x^2+\|x\|$ corresponds to the radial equation $-u''(r)+ r u(r)=e_0 u(r)$ with $u(0)=u(\infty)=0$ and is the negative of the first zero of the Airy function, $e_0=2.3381\ldots$\@
By scaling, the ground state of $\alpha p_x^2+\beta \|x \|$, with $\alpha>0$ and $\beta>0$ is 
$\alpha^{1/3}\beta^{2/3} e_0$. Thus the lowest eigenvalue of $H'$ is
\begin{equation}\label{eq:boundE1}
E'=e_0\left[ M^{-1/3}+1+(4\mu)^{-1/3}\right]~,
\end{equation}
with $\mu=M/(1+M)$. By comparison, the threshold of $(QQ\bar{q}\bar{q})$ is made of two identical $(Q\bar{q})$ mesons, each governed by the Hamiltonian $h=p^2/(2 \mu) + \| r \|$, where $p$ is conjugate to the quark--antiquark separation $r$. Thus the threshold energy is
\begin{equation}\label{eq:threshold}
E_\text{th}=2\, e_0(2 \mu)^{-1/3}~,
\end{equation}
and it is easily seen that $E'>E_\text{th}$ for any value of the quark-to-antiquark mass ratio $M$, i.e., the bound (\ref{eq:boundU1}) cannot demonstrate binding.

A better bound will be proved below. If there is a genuine Steiner tree%
\footnote{This will be made more precise in the proof given in Appendix}
 linking the quarks to the antiquarks, then 
\begin{equation}\label{eq:boundV_4}
V_4\le \frac{\sqrt3}{2}\left(\| x\| +\| y\| \right)+ \| z \| ~.
\end{equation}
But if $V_4$ is not associated to a genuine Steiner tree,  this inequality is often violated. Consider for instance  a rectangular configuration with $||v_1v_2||=||v_3v_4|| \gg ||v_1v_3||=||v_2v_4||$ (in this case the mathematical Steiner tree problem would require a Steiner point linking $v_1$ and $v_3$, another Steiner point linking $v_2$ and $v_4$, but the corresponding fluxes are not permitted by the color coupling in QCD), then $\| z\| \sim 0$ and $V_4\sim \|x\|+\| y\|$, so (\ref{eq:boundV_4}) does not hold.

\begin{figure}[!htbc]
\begin{minipage}{.55\textwidth}
\includegraphics[width=.95\textwidth]{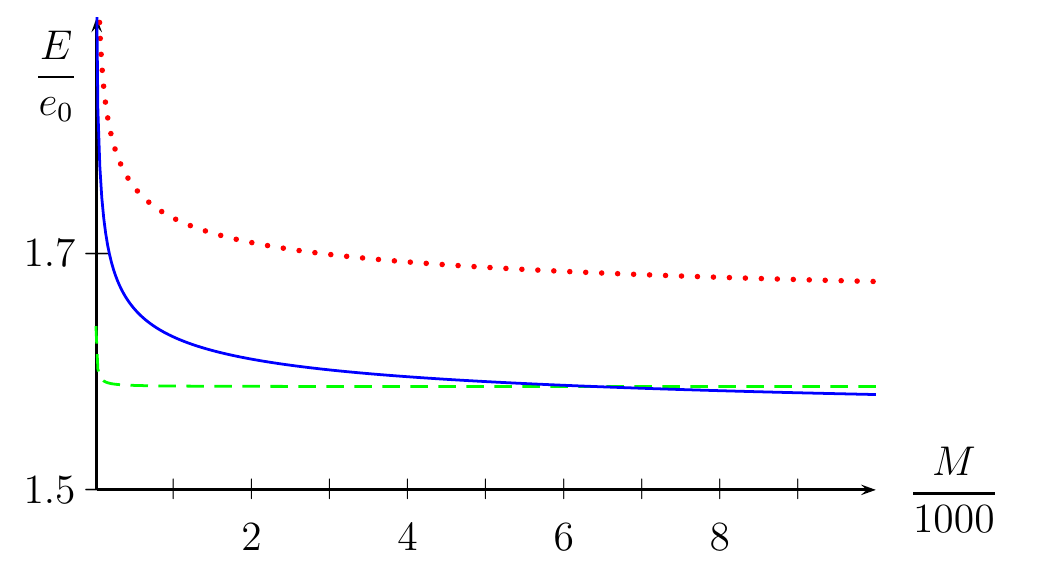}
\end{minipage}
\begin{minipage}{.44\textwidth}
\caption{\label{fig:bounds}  Simple bound $E'$ (Eq.~(\ref{eq:boundE1}), dotted line) and improved upper bound $E''$ (Eq.~(\ref{eq:boundE2}),solid line) on the tetraquark ground-state energy as a function of quark-to-antiquark mass ratio $M$. Also shown is the threshold energy $E_\text{th}$ (Eq.~\ref{eq:threshold}), dashed line). The energies are in units of $e_0$, the ground state of $-\Delta + \| r \|$.}
\end{minipage}
\end{figure}

However, it will be shown that
\begin{equation}\label{eq:boundU2}
U\le \frac{\sqrt3}{2}\left(\| x\| +\| y\| \right) + \| z \| ~,
\end{equation}
for any configuration of the quarks and antiquarks, i.e., for any $x$, $y$ and $z$.
Then the ground state of $H$ is bounded as
\begin{equation}\label{eq:boundE2}
E<E''=e_0\left[ 
\left(\frac{3}{4}\right)^{1/3}\!\left(M^{-1/3}+1\right)+(4\mu)^{-1/3}
\right]~,
\end{equation}
As shown in Fig.~\ref{fig:bounds}, this bound $E''$ significantly improves the previous one, $E'$. 
It is easily seen than $E''$ becomes smaller than $E_\text{th}$ for very large values of the mass ratio, more precisely for $M>6402$, and thus that the tetraquark  is bound at least in this range of $M$.
The numerical estimate of \cite{Vijande:2007ix} actually indicates stability for all values of $M$, even $M=1$.

To summarize, we obtained an analytic upper bound on the ground state energy of tetraquarks systems with two units of open flavor, $(QQ\bar{q}\bar{q})$, using a model of linear confinement inspired by the strong-coupling regime of QCD. The key is an inequality on the length of a Steiner tree with four terminals.
The bound confirms a recent numerical investigation, in which this potential was shown to bind these tetraquarks below the threshold for dissociation into two mesons. It remains to investigate whether this stability survives  refinements in the dynamics, such as short range corrections, spin-dependent forces, etc.

It is our intention to extend this investigation to the case of the pentaquark (one antiquark and four quarks) and hexaquark configurations (six quarks), which have been much debated in recent years.

\paragraph*{Acknowledgements}
We thank Emmanuelle Vernay and Sylvie Flores for their help in collecting the bibliography, and also J.-C.~Angles d'Auriac, M.~Asghar, A.~Valcarce and J.~Vijande for enjoyable discussions on this problem, as well as Dave Eberly for useful correspondence.
\subsection*{Appendix: Results on the Steiner problem}
Before deriving Eq.~(\ref{eq:boundU2}), let us review some basic properties of the string potentials $V_3$ and $V_4$.

\begin{figure}[htbc]
\begin{center}
\subfigure[]{\label{Fig:Three-a}
\includegraphics[]{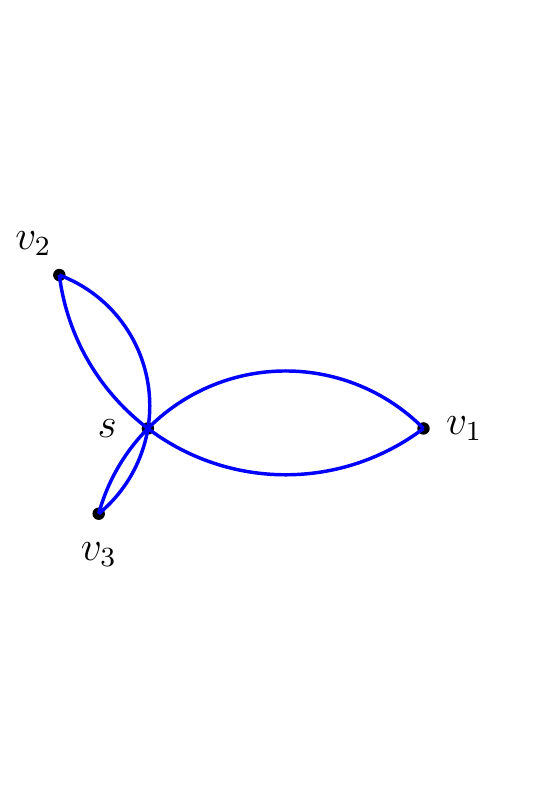}}
\hskip 1cm
\subfigure[ ]{\label{Fig:Three-b}
\includegraphics[]{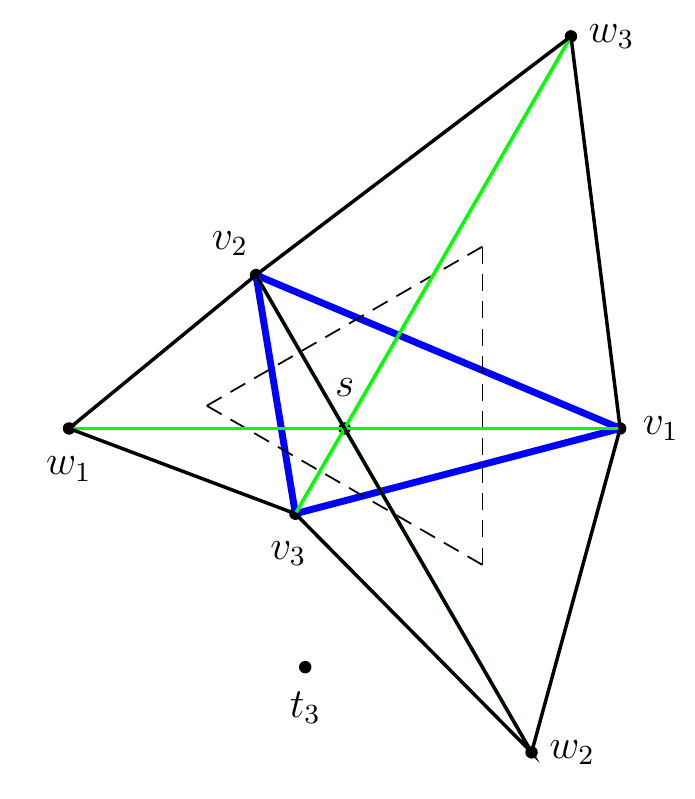}}
\end{center}
\caption{\label{Fig:Three} Left: the junction lies on each arc from which a side is seen under $120^\circ$. Right: the three-terminal Steiner problem as a side product of Napoleon's theorem}
\end{figure}

\subparagraph{Three terminals}
The three-point problem is very much documented in textbooks \cite{Coxeter:100457,Courant:397053,Hwang:1992,Proemel:2002,0842.90116}.
Let $v_1v_2v_3$ be the triangle, with side lengths $a_1=||v_2v_3||$, \dots\ and angles $\alpha_1=\angle{v_1v_2v_3}$, etc.\@
The problem of finding a path of minimal length $\| sv_1\|+\| sv_2\| +\| sv_3\|$ linking the three vertices has been solved by Fermat and Torricelli. See, e.g., \cite{Coxeter:100457}. The result is the following: if one of the angles, say $\alpha_1$, is larger than $120^\circ$,  $s$ coincides with $v_1$, otherwise each side of the triangle is seen from  $s$ with an angle of  $120^\circ$. The Steiner point $s$ is thus at the intersection of three arcs of  circles, see Fig.~\ref{Fig:Three-a}.

The three-terminal problem is also linked to Napoleon's theorem, which states that if one draws 
 external equilateral triangles on each side, $v_1v_2w_3$, $v_1v_3w_1$ and $v_1v_1w_2$, the centers of these triangles  form an \emph{equilateral} triangle (dashed lines in Fig.~\ref{Fig:Three-b}), a nice example of symmetry restoration. The junction $s$ is just the intersection of $v_1w_1$, $v_1w_2$ and $v_3$. Note that $\| sv_1\| +\| sv_2\| =\| sw_3\|$, and similar relations, and thus the potential is simply $V_3=\| v_1w_1\| =\| v_2 w_2 \|=\| v_3 w_3 \|$.

The point $w_3$ and its symmetric with respect to $v_1v_2$, $t_3$ form the toroidal domain associated to the subset $\{v_1,v_2\}$.  The length of the minimal Steiner tree is the maximal distance between $v_3$ and the domain $\{w_3,t_3\}$.
 
From the above properties, one can estimate the string potential in a closed form. If $\alpha_1\ge 120^\circ$, then $V_3=a_2+a_3$, and similarly for large $\alpha_2$ or $\alpha_3$. Otherwise, $V_3=\sum \ell_i$, where $\ell_i=\| sv_i\|$. From $\hat{s}=120^\circ$ in the triangle $sv_2v_3$, $a_1^2=\ell_2^2+\ell_3^2+\ell_2\ell_3$, and by summation
\begin{equation}\label{eq:sumJAs}
2(\ell_1^2+\ell_2^2+\ell_3^2)+(\ell_1\ell_2+\ell_2\ell_3+\ell_3\ell_1)=a_1^2+a_2^2+a_3^2~.
\end{equation}
Now, $\ell_1\ell_2$ being four times the area of the $sv_1v_2$ triangle, the second term in the above equation is four times the whole area of $v_1v_2v_3$, which is given by the Henon theorem. 
%
Altogether, in the case of a genuine Steiner tree \cite{Takahashi:2000te}
\begin{multline}\label{eq:Jap}
V_3=\ell_1+\ell_2+\ell_3=
\sqrt{a_1^2+a_2^2+a_3^2+ \sqrt{3(a_1+a_2+a_3)(a_1+a_2-a_3)(a_1-a_2+a_3)(-a_1+a_2+a_3)}}~,
\end{multline}
which can be computed quickly.
\subparagraph{The planar tetraquark problem.}
%



\begin{figure}[!ht]
\begin{center}
\subfigure[\ Large angle in $v_1$ ]{\label{Fig:sp4-a}
\raisebox{1cm}{\includegraphics[]{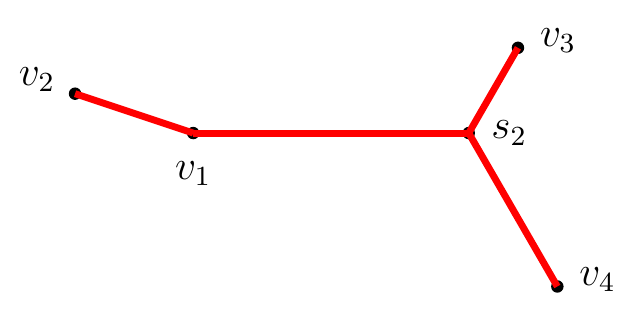}}}
\hspace*{1cm}
\subfigure[\ Elongated rectangle]{\label{Fig:sp4-b} \includegraphics[]{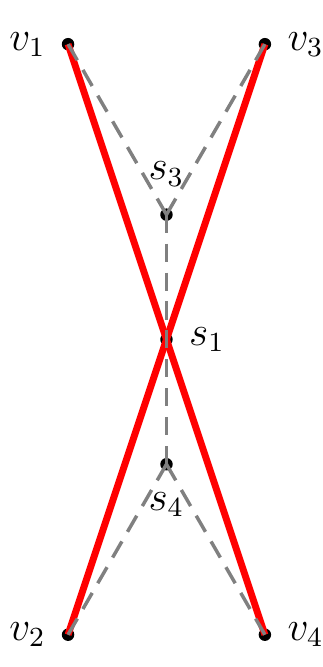}}
\end{center}
\caption{\label{Fig:sp4} Examples of special configurations. Left: one junction coincides with $v_1$. Right: the two junctions merge (the dotted gray line corresponds to the Steiner tree if the four points $v_i$ play the same role, unlike the tetraquark problem with quarks and antiquarks having conjugate colors. )}
\end{figure}

For the four-point problem, there are many special cases, which can be treated by inspection. If, for instance the quark $v_2$ is on the back of $v_1$, as in Fig.~\ref{Fig:sp4-a}, the problem reduces to the Steiner problem for $\{v_1,v_3,v_4\}$. Another special case is shown in Fig.~\ref{Fig:sp4-b}, where the quarks are close to the antiquarks. For the standard Steiner problem of geometry, the solution would correspond to the Steiner tree shown as a dotted line, with a Steiner point $s_3$ linked to $v_1$ and $v_3$ and another one, $s_4$, linked to $v_2$ and $v_4$.  This is not allowed by the different color properties of quarks and antiquarks, hence our best tree, shown as a solid line, has only one junction.  But in estimating the potential $U$ of Eq.~(\ref{tetra-pot}) for this configuration, the minimum is the flip--flop term $d_{13}+d_{24}$.

%

\begin{figure}[!tcb]
\begin{center}
\includegraphics[]{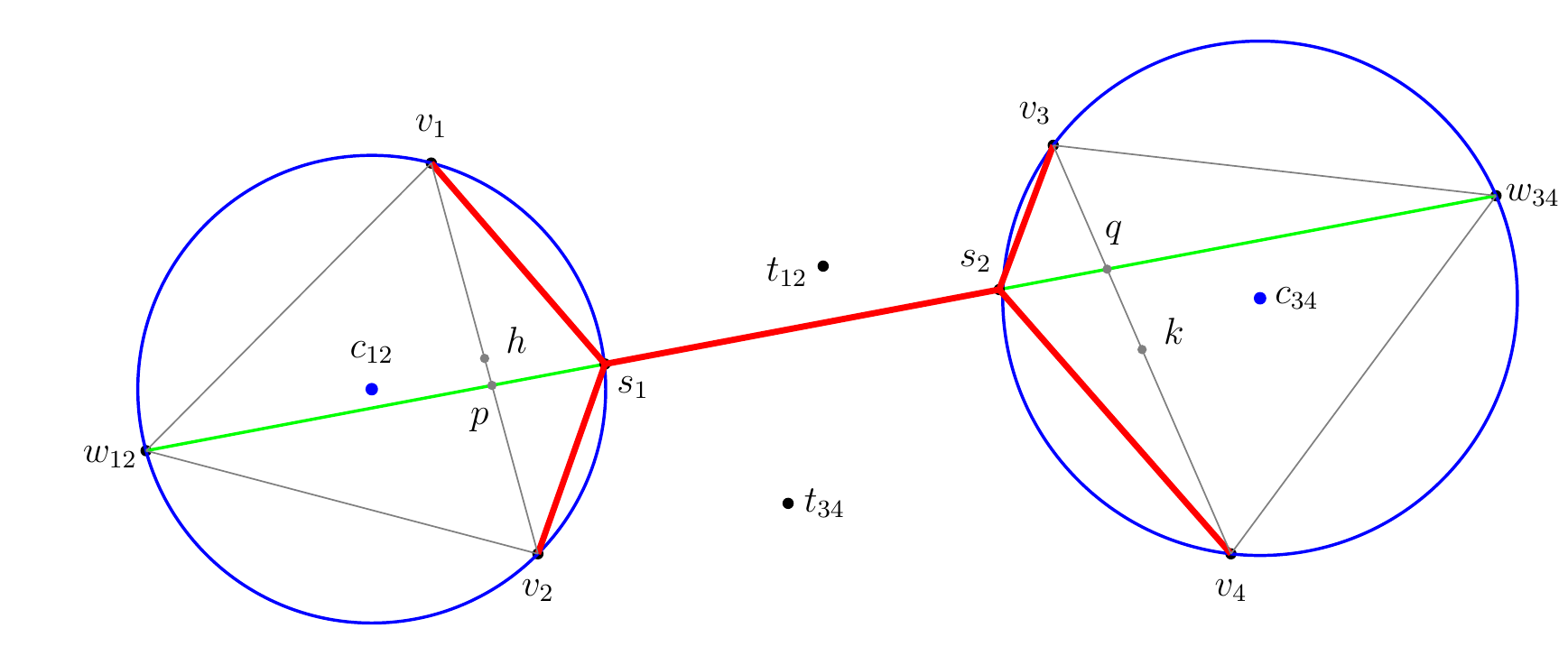}
\end{center}
\caption{\label{Fig:4plan}Construction of the minimal string in the planar case.}
\end{figure}

Let us turn to the case of a genuine  Steiner tree $(v_1v_2)s_1s_2(v_3v_4)$ as in Fig.~\ref{Fig:4plan}.
The string of Fig.~\ref{fig:bar-tetra-Steiner} is minimized with respect to $s_1$ and $s_2$. Hence for fixed $s_2$, it assumes the Fermat--Torricelli minimization of $v_1v_2s_1$, a well-known iteration property of Steiner trees. Hence $\angle{v_1s_1v_2}=120^\circ$ and $v_1v_2$ is the bissector of $\angle{v_1s_1v_2}$ and passes through the point $w_{12}$ which completes an equilateral  triangle $v_1v_2w_{12}$ in the quark sector.
Similarly, it also passes through $w_{34}$ which makes $v_3v_4w_{34}$ equilateral in the antiquark sector.

The junction points $s_1$ and $s_2$ are just the other intersections of the straight line $w_{12}w_{34}$ with the circumcircles of $v_1v_2w_{12}$ and $v_3v_4w_{34}$, as shown in Fig.~\ref{Fig:4plan}. There is a possible ambiguity about on which side $s_1$ or $s_2$ should be, but this is easily solved by the requirement that the total length of the string is minimum.  Crucial is the observation  that $V=\| w_{12}w_{34}\|$,  so that the determination of the Steiner points $s_1$ and $s_2$ is not required to compute $V_4$.

A variant is that is $t_{12}$ is the symmetric of $w_{12}$ with respect to $v_1v_2$, the set $\{w_{12}t_{12}\}$ is the toroidal domain associated to the quarks, and similarly $\{w_{34} t_{34}\}$ for the antiquarks, the length of the Steiner trees is the maximal distance between these two sets.

This construction, which is a special case of the Melzak's algorithm \cite{0101.13201}, leads to a very easy computation. If each vector $v_i$ is identified with its affix (complex number) $v_i$, etc.,   then those of $w_{12}$ and $w_{34}$ are easily deduced, for instance $w_{12}=-j^2 v_1-j v_2$ or $-j v_1-j^2 v_2$ (depending on which side is $w_{12}$), if one uses the familiar root of unity $j=\exp(2i\pi/3)$. Once $w_{12}$ and $w_{34}$ are determined, $V=\| w_{12} w_{34} \|$. If one wishes to locate the Steiner points, it is sufficient to remark that $w_{12}s_2.w_{12}w_{34}=\| w_{12}c_{34}\|^2-r_{34}^2$ and $w_{34}s_1.w_{34}w_{12}=\| w_{34}c_{12}\|^2-r_{12}^2$, where $c_{12}$ is the center of the circle $v_1v_2w_{12}$ and $r_{12}=d_{12} \sqrt{3}/2$ its radius and $c_{34}$ and $r_{34}$ are defined similarly in the antiquark sector.
\subparagraph{The spatial tetraquark problem}
In general, the four constituents do not belong to the same plane. The minimum is achieved for $v_1v_2s_1s_2$ coplanar, and $v_3v_4s_1s_2$ also coplanar, but in a different plane. The toroidal domain to which the point $w_{12}$ belongs is the  Melzak circle, of axis $v_1v_2$ and radius $r_{12}=\|v_1v_2\| \sqrt{3}/2$, and similarly for $w_{34}$ in the antiquark sector. The straight line $w_{12}w_{34}$ has to intersect  these two circles as well as the lines $v_1v_2$ and $v_3v_4$. The problem consists of constructing such a straight line.

The reasoning can be made on Fig.~\ref{Fig:4plan}, if one imagines that $v_3v_4s_2$ is not coplanar to $v_1v_2s_1$. As stressed in \cite{Rubinstein:2002}, the key is to determine $p$ and $q$, the intersections of $s_1s_2$ with $v_1v_2$ and $v_3v_4$, respectively. In this paper, the following coupled equations are derived
\begin{equation}\label{eq:Rubinstein}
x_p={m\sqrt{h^2+x_q^2\sin^2\phi}+r_{12} v\cos\phi\over r_{34}+ \sqrt{h^2+x_q^2\sin^2\phi}}~,
\quad
x_q={n\sqrt{h^2+x_p^2\sin^2\phi}+r_{34} v\cos\phi\over r_{12}+ \sqrt{h^2+x_p^2\sin^2\phi}}~,
\end{equation}
for the abscissa $x_p$ of $p$ along $v_1v_2$ and $x_q$ of $q$ along $v_3v_4$. These abscissas  are from the common perpendicular $uv$ to $v_1v_2$ and $v_3v_4$ ($u\in v_1v_2$and $v\in v_3v_4$), with $\| uv \|=h$, $\| uh\| =m$ and $\| vk \|=n$, where $h$ is the middle of $v_1v_2$ and $k$ that of $v_3v_4$.
The equations (\ref{eq:Rubinstein}) can be solved by iterations, with remarkably fast convergence. Once $x_p$ and $x_q$, i.e., $p$ and $q$, are determined, the Steiner points are determined  by imposing they are on the circles $v_1v_2w_{12}$ and $v_3v_4w_{34}$, respectively. For instance, if $s_1=p+t(q-p)$, $t$ obeys a second order equation\footnote{There is a misprint in \cite{Rubinstein:2002} which propagated in the numerical calculation given as an example.}. 

If one is interested only in the length of the Steiner tree and not in the position of the Steiner points,  an alternative formalism consists of locating $p$ through $p=h+x\,(v_2-h)$ and $q=k+y\,(v_4-k)$. With this notation,  the length of the tree is simply
\begin{equation}\label{eq:tobemin}
V_4=\min_{x,y}\left[ \| pq\|+{r_{ab}\over\sqrt{3}}\sqrt{3+x^2}+{r_{cd}\over\sqrt{3}}\sqrt{3+y^2}\right]~,
\end{equation}
which is easily minimized by varying $x$ and $y$. The minimisation is equivalent to solving the coupled equations
\begin{equation}\label{eq:ourcoupl}
x=\sqrt{3+x^2} {{v_1v_2}. {pq}\over \| v_1v_2\| \,\| pq\| }~,
\quad
y=\sqrt{3+y^2} {{v_3v_4}.{qp}\over \| v_3v_4\| \,  \| pq\| }~,
\end{equation}
which expresses that $w_{12}$, $p$, $s_1$, $s_2$, $q$ and $w_{34}$ are collinear. These equations are easily solved by iteration or any other means. 

We believe that, besides checking the particular cases with large angles or a single Steiner point, the fastest computation of the connected four-quark potential consists of minimising 
(\ref{eq:tobemin}) or solving (\ref{eq:ourcoupl}). We expect a dramatic improvement in computing time from the above algorithm.

However, it is aesthetically appealing to attempt a further reduction of the number of variables to be determined numerically, and to provide an almost analytic estimate of the interaction as a function of the coordinates of the quarks and antiquarks. Finding $V_4=\|w_{12}w_{34}\|$, the maximal distance between the Melzak circles $\mathcal{C}_{12}$ and $\mathcal{C}_{34}$, is very similar to the problem of the minimal distance beween two circles in space, as addressed e.g., in \cite{106259,Eberly}. Neff \cite{106259} has shown that with the help of Lagrange multipliers and Gr{\"o}bner type of elimination performed by computer-algebra sofware,
the squared stationary distance $V_4^2$ obeys an eighth-order polynomial equation whose coefficients are rational functions of the coordinates of $v_1$, $v_2$, $v_3$ and $v_4$.
 
Eberly \cite{Eberly} showed that if $m$ is associated to an angle $\theta$ along $\mathcal{C}_{12}$, and $n$ to $\phi$ along $\mathcal{C}_{34}$, then imposing $\| mn \| ^2$ to be stationary, results in two equations of the type
\begin{equation}\label{eq:eberly}
\alpha_i\cos\theta +\beta_i\sin\theta+\gamma_i=0~,\qquad i=1,2~,
\end{equation}
where $\alpha_i$, $\beta_i$ and $\gamma_i$ contain constants and terms linear in $\cos\phi$ and $\sin\phi$. Solving (\ref{eq:eberly}) as two linear equations, as if $\cos\theta$ and $\sin\theta$ were independent, and then imposing $\cos\theta^2+\sin\theta^2=1$ gives an equation for $\cos\phi$ and $\sin\phi$, which is transformed into an 8$^{\rm  th}$ order equation in $\cos\phi$.

It is slightly faster to rewrite (\ref{eq:eberly}) using $t=\tan(\theta/2)$ and $u=\tan(\phi/2)$ as
\begin{equation}\label{eq:eberly1}
\delta_i t^2+ \eta_i t+\epsilon_i=0~,\qquad i=1,2~,
\end{equation}
where the coefficients are quadratic in $u$. The compatibility of two such equations is simply
\begin{equation}\label{eq:eberly2}
W(\delta,\eta)W(\eta,\epsilon)=W(\delta,\epsilon)^2~,\qquad
W(x,y)=x_1y_2-x_2y_1~,
\end{equation}
and is directly a polynomial in $u$, of order 8.
\begin{figure}[!!htbc]
\begin{minipage}{.4\textwidth}
\begin{flushleft}
\includegraphics[width=.9\textwidth]{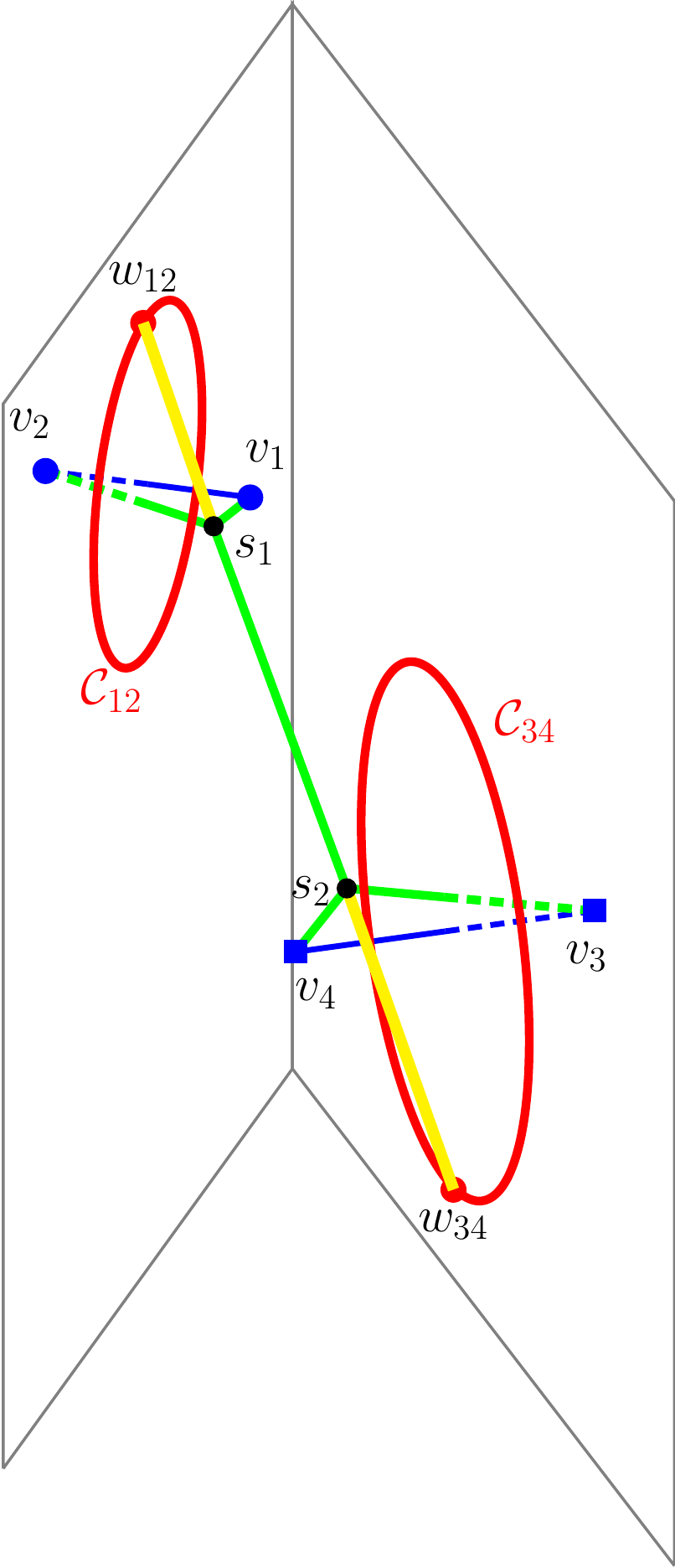}
\end{flushleft}
\end{minipage}
\begin{minipage}{.59\textwidth}
\caption{\label{fig:3d} The confining potential $V_4$ for the tetraquark system $(v_1v_2v_3v_4)$ is the minimal length of the tree $\| v_1s_1\| +\| v_2s_1\| +\| s_1s_2\| + \| s_2v_3\| + \| s_2v_4\|$ when $s_1$ and $s_2$ are varied. It is also the maximal distances between the circles $\mathcal{C}_{12}$ and $\mathcal{C}_{34}$, i.e., the distance $w_{12}w_{34}$. The Melzak circle $\mathcal{C}_{12}$ is centered at the middle of $v_1$ and $v_2$, has $v_1v_2$ as axis and a radius $\| v_1v_2\| \sqrt{3}/2$, and $\mathcal{C}_{34}$ has analogous properties in the antiquark sector.}
\end{minipage}
\end{figure}

\subparagraph{Proof of the inequality  (\ref{eq:boundU2})}
%
 
 If we have a positively oriented edge from $s_1$ to $s_2$, i.e., the Steiner tree is non degenerate, then we have $$V_4 \le (\|x\|+\|y\|) {\sqrt 3} /2 + \|z\| =B$$ using Melzak circles. 
 
However the  bound required is for $U=$ min $\{ d_{13}+d_{24}, d_{14}+d_{23},V_4\}$. So we want to confirm that  $$U \le (\|x\|+\|y\|) {\sqrt 3} /2 + \|z\| =B $$ is valid, regardless of whether $V_4$ is a degenerate or non degenerate Steiner tree. 
 
 We follow the variational method introduced in \cite{0734.05040}. The problem is formulated as a global optimisation problem as follows;
 
Define $L$ as the length of the {\it formal} Steiner tree spanned by the four vertices. This length is obtained from the distance between the farthest points on the two Melzak circles. In terms of the usual Steiner tree components, $L=\|v_1 s_1\|+\|v_2 s_1\|\pm \|s_1 s_2\|+ \|v_3 s_2\|+\|v_4 s_2\|$).  We get the positive sign for $\|s_1 s_2\|$ if there is a real Steiner tree. On the other hand, if the Steiner vertices have interchanged position, so that on the line between the two farthest Melzak points, $s_2$ is closer to the Melzak point for $v_1,v_2$ than $s_1$, then we have the negative sign for $\|s_1 s_2\|$. 
So we can construct a formal tree on the six vertices $v_1,v_2,v_3,v_4,s_1,s_2$ where the edge joining the two Steiner vertices is `negatively oriented'. 

Now it is easy to see that $L \le (\|x\|+\|y\|) {\sqrt 3} /2 + \|z\| $. So if $V=L$ then the desired inequality follows trivially. So we only need to consider the situation where $L<V$, i.e the Steiner tree is formal rather than a real Steiner tree. Now by the inequality above, if either of $d_{13}+d_{24}, d_{14}+d_{23}$ is not larger than $L$, then clearly the required inequality follows. So we only need to consider the case when $d_{13}+d_{24} >L$ and  $d_{14}+d_{23}>L$. 

We can parametrise the points $v_1$, $v_2$, $v_3$, $v_4$ by the numbers  $\|v_1 s_1\|$, $\|v_2 s_1\|$,  
 $\pm \|s_1 s_2\|$, $\|v_3 s_2\|$, $\|v_4 s_2\|$. (It is easy to see that these four points are determined up to rotation, translation by five parameters.) By rescaling, we can assume that the sum of these five numbers is $1$, without loss of generality for the inequality. It is easy to see that all the numbers are then bounded so the domain becomes compact. So we seek a maximum of the ratio of $R=\min\{ d_{13}+d_{24}, d_{14}+d_{23}\}$ and $(\|x\|+\|y\|) {\sqrt 3} /2 + \|z\| =B $ over this domain. 

Now suppose that we rotate the triangles $v_1 v_2 v_3$ and $v_1 v_2 v_4$ around an axis line through $v_1 v_2$. Clearly we can think of one triangle as being fixed and the other as moving relative to the first one. The quantity $R$ does not change by this rotation, but obviously $B$ does. Hence a maximum of the ratio $R/B$ corresponds to a minimum for $B$ under such a rotation. 

Now an elementary argument shows that such a minimum for $B$ occurs for the configuration being planar, i.e when the vertex $v_4$ moves into the plane of $v_1,v_2,v_3$. Now assume that some initial configuration satisfies $R/B>1$ and the Steiner tree is formal rather than real. As the triangle $v_1 v_2 v_4$ rotates around an axis line through $v_1 v_2$, it is easy to see that the two Melzak circles move apart. At some intermediate point, if they cross, then we find that the Steiner tree changes from being formal to being real. At this intermediate point, it is trivial to see that $R/B<1$. But this is impossible, since we have initially $R/B>1$ and $R/B$ is increasing, since $B$ is decreasing and $R$ is fixed. 

On the other hand, if the Melzak circles never intersect, then this must be true for the planar configuration. So we would have such a configuration for which the Steiner tree is still formal but $R/B>1$. It is elementary to prove that this is impossible. So this completes the argument. 
%

\end{document}